\documentclass[superscriptaddress,showpacs,twocolumn,nofootinbib,aps,prl]{revtex4-2}
\usepackage{comment}
\usepackage{fontawesome5}
\usepackage{float}
\usepackage[percent]{overpic}
\usepackage[normalem]{ulem}
\usepackage{graphicx}
\usepackage{color}
\usepackage{xcolor}
\usepackage{verbatim}
\usepackage{subfigure}
\usepackage{amssymb}
\usepackage{amsmath}
\usepackage{comment}
\usepackage{chngcntr}
\usepackage{appendix}
\usepackage{lipsum}
\usepackage{dsfont}
\usepackage{xfrac}
\usepackage{xr}
\usepackage[makeroom]{cancel}
\newcommand{\dd}{\text{d}}

\newcommand{\ee}{\text{e}}

\newcommand{\p}{\partial}

\newcommand{\subfigref}[2]{\hyperref[#1]{\ref*{#1}#2}} 

\usepackage{amsthm}
\usepackage{bm}
\usepackage[hang,flushmargin]{footmisc}
\usepackage[colorlinks = true,linkcolor = blue,urlcolor = blue,citecolor = blue,anchorcolor = blue]{hyperref}
\definecolor{greencross}{RGB}{83, 170, 46}

\begin{document}

\title{The Mpemba effect likes to hit a wall}
\author{Yue Liu}
\affiliation{Center for Gravitational Physics and Quantum Information, Yukawa Institute for Theoretical Physics, Kyoto University, Kitashirakawa Oiwakecho, Sakyo-ku, Kyoto 606-8502, Japan}

\author{Tan Van Vu}
\affiliation{Center for Gravitational Physics and Quantum Information, Yukawa Institute for Theoretical Physics, Kyoto University, Kitashirakawa Oiwakecho, Sakyo-ku, Kyoto 606-8502, Japan}

\author{Raphaël Chétrite}

\affiliation{CNRS Laboratoire Ypatia des Sciences Mathématiques (LYSM), Piazzale Aldo Moro 5, 00185 Rome, Italy}


\author{Frédéric van Wijland}
\affiliation{Laboratoire Mati\`ere et Syst\`emes Complexes (MSC), Université Paris Cité  \& CNRS (UMR 7057), 75013 Paris, France}
\affiliation{Center for Gravitational Physics and Quantum Information, Yukawa Institute for Theoretical Physics, Kyoto University, Kitashirakawa Oiwakecho, Sakyo-ku, Kyoto 606-8502, Japan}

\author{Hisao Hayakawa}
\affiliation{Center for Gravitational Physics and Quantum Information, Yukawa Institute for Theoretical Physics, Kyoto University, Kitashirakawa Oiwakecho, Sakyo-ku, Kyoto 606-8502, Japan}
\begin{abstract}
The original Mpemba effect refers to the counterintuitive possibility that a system prepared at a higher temperature relaxes to a colder equilibrium faster than one prepared at a lower temperature. In microscopic classical realizations, this phenomenon has been studied for overdamped particles in external potentials and observed experimentally. We show that, for a particle in an asymmetric polynomial double-well potential, the one-dimensional Mpemba effect is controlled not by the double-well structure or metastability itself, but by the high-energy confinement: a hard boundary, or more generally a sufficiently steep confining tail, on the shallow side can induce the required population inversion. For the polynomial potentials considered here, the effect disappears when both hard boundaries are taken to infinity, whereas sufficiently asymmetric soft tails can play the same role as walls. We identify the underlying mechanism as a redistribution of the initial equilibrium population when it begins to probe the high-energy tails of the potential, a picture that also extends to single-well and symmetric double-well potentials.

\end{abstract}

\maketitle

\vspace{0.1in}

Since the seminal report by Mpemba and Osborne~\cite{mpemba1969cool}, it has been recognized that a system initially prepared at a higher temperature may relax to equilibrium faster than one prepared at a lower temperature when both are quenched into a cold bath. 
This counterintuitive phenomenon, now broadly referred to as the Mpemba effect, has become a paradigmatic example of anomalous relaxation.
While its existence in macroscopic systems, often phrased as faster freezing from hotter initial conditions, remains under active debate~\cite{burridge2016questioning,katz2017reply}, compelling experimental evidence has established its presence in controlled settings. In particular, the experiment of Ref.~\cite{kumar2020exponentially}, involving a colloidal particle in a tunable one-dimensional (1D) optical potential, demonstrated such anomalous relaxation by varying the initial bath temperature.
These findings have stimulated extensive investigations of the Mpemba effect across a wide range of systems, spanning colloidal~\cite{kumar2020exponentially, chetrite2021metastable,Kumar22,Biswas23a,Gianluca2023b}, classical~\cite{lu2017nonequilibrium,Santos17,klich2019mpemba,Baity-Jesi19,Mompo20,Biswas20,Yang20,Gonzalez2021,Biswas21,Takada21a,walker2021anomalous,Busiello21,Holtzman22,Gianluca2023a,Biswas23,Biswas24,Santos24,Ohga2024,Tan2025,Nava2025,Nava20252,Teza2026}, and quantum regimes~\cite{Carollo21, Ares23,Chatterjee_2023, Chatterjee2024,Rylands2024, Moroder2024,Yamashika2024,Liu24,Wang24,Nava24,Longhi24,Joshi2024,Shapira2024,Zhang2025, Yamashika2025,Strachan2025,Turkeshi2025,Bao2025,Yu2025,Beato2026,Yamashika2026}.

\begin{figure}[t]
\begin{center}
\includegraphics[width=1\linewidth]{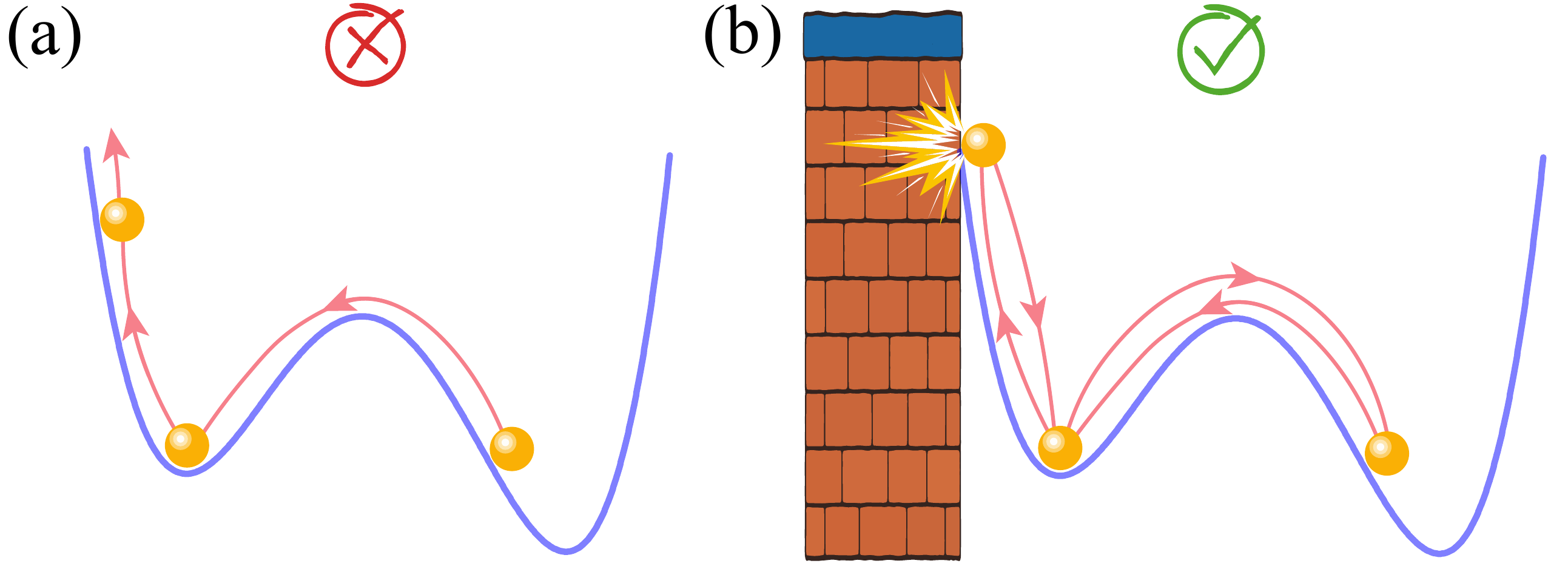}
\caption{The presence (\textcolor{greencross}{\faCheckCircle[regular]}) or absence (\textcolor{red}{\faTimesCircle[regular]}) of the Mpemba effect is shown for an asymmetric potential without wall (a) and with a left wall (b). The presence of a wall on the shallow side is a necessary condition for the existence of the Mpemba effect. The blue solid lines represent the potential, and the red arrows that begin from the right well indicate the direction of population transfer as the initial temperature is increased.
}
\label{fig:cartoon}
\end{center}
\end{figure}

The idea of Ref.~\cite{lu2017nonequilibrium}, experimentally investigated in~\cite{kumar2020exponentially,tian2025experimental}, is to use a double-well potential to mimic the metastability underlying a macroscopic first-order transition. 
We restrict our interest to the behavior of a particle confined in an asymmetric double-well potential in this Letter.
We will generalize this argument to general 1D potentials (symmetric, single well, with or without walls) in a longer version~\cite{pre-us}.
Although some works on the Mpemba effect consider nonequilibrium initial conditions~\cite{Santos17,Mompo20,Biswas20,Yang20,Gonzalez2021,Biswas21,Takada21a,Carollo21,Degunther2022,Biswas23,Ares23,Chatterjee_2023, Chatterjee2024,Rylands2024, Moroder2024,Yamashika2024,Liu24,Wang24,Nava24,Longhi24,Joshi2024,Shapira2024,Zhang2025, Yamashika2025,Strachan2025,Turkeshi2025,Bao2025,Beato2026,Yamashika2026}, we choose to focus on the original setup in which we compare the dynamics from two initial equilibrium conditions at different temperatures~\cite{mpemba1969cool,lu2017nonequilibrium,klich2019mpemba,Baity-Jesi19,kumar2020exponentially,walker2021anomalous,chetrite2021metastable,Kumar22,Holtzman22,Biswas23a,Gianluca2023b,Gianluca2023a,Malhotra2024,Biswas24,Santos24,Ohga2024,Tan2025,Nava2025,Nava20252,tian2025experimental,Antonov2026}. 

The physical gist of a double-well potential is to introduce a timescale (which can be made arbitrarily large as the bath temperature is lowered) during which the anomalous relaxation effect of interest can be properly observed. 
In this Letter, for analytical simplicity, we first consider 1D double-well potentials confined by two hard walls, whose positions can subsequently be taken to infinity. Throughout this work, however, the term ``wall'' is used in the broader sense of a hard boundary or a sufficiently steep confining tail. The extension to soft confinement is discussed in Ref.~\cite{pre-us} and in the Supplemental Material (SM)~\cite{SM}.
Many of the references address this setting~\cite{kumar2020exponentially,Bechhoefer2021, chetrite2021metastable,walker2021anomalous,Biswas23a,Malhotra2024,tian2025experimental}, each emphasizing different physical mechanisms underlying anomalous relaxation, including the role of the energy barrier separating the two potential wells. 
These works adopt various realizations, ranging from smooth to sawtooth or square double-well potentials. 
However, mutual comparison is often cumbersome, simply because the parameter space to explore, location of the wells, widths, curvatures, well depths, barrier height, slopes, is huge. 
Our work rationalizes these various physical ingredients.

We argue that the key physical mechanism controlling the Mpemba effect in an asymmetric polynomial double-well potential has not been fully isolated. In the hard-wall setting studied analytically below, a confining boundary on the shallow side is necessary for the effect: when that boundary is removed to infinity, the Mpemba temperature is pushed to infinity and the effect disappears. The same mechanism can be generated without a literal hard boundary when the shallow-side tail is sufficiently steeper than the opposite tail~\cite{pre-us,SM}. Thus, in what follows, ``wall-driven'' refers to confinement by either a hard boundary or an effectively wall-like steep tail. Such confinement is necessary within the class considered here but is not sufficient, because a suitably placed wall on the deep side can suppress the population inversion. We first establish these statements analytically at low bath temperature and then test their persistence numerically at finite bath temperature.
We then fill in the intermediate range of bath temperatures with numerical simulations. We synthesize our findings by proposing a mechanism based on a detailed analysis of the response of the initial equilibrium population with respect to initial temperature change. Our findings are summarized in the cartoon of Fig.~\ref{fig:cartoon}.

The setting of our study is a 1D overdamped Langevin equation describing a particle with position $x(t)$ evolving in a potential landscape $V(x)$ with unit mobility
\begin{equation}\label{eq:Langevin}
    \frac{\dd x(t)}{\dd t}=-V'(x(t))+\sqrt{2T}\eta(t),
\end{equation}
where $\eta$ is a zero-mean Gaussian white noise with correlations $\langle\eta(t)\eta(t')\rangle=\delta(t-t')$, $T=\beta^{-1}$ is the bath temperature (Boltzmann's constant $k_\text{B}$ is set to unity), and the potential $V$ becomes infinity at $x=-L_-$ and $x=L_+$. 
The Fokker-Planck evolution equation for the probability density $p(x,t)$ to find the particle at position $x$ at time $t$ reads
\begin{equation}\label{eq:FP}
    \p_t p={\mathbb W}p=\p_x(V'p)+T\p_x^2 p.
\end{equation}
For an initial condition $p(x,0)$, the solution at time $t$ is obtained as
\begin{equation}
p(x,t)=\sum_{n\ge 1}\ee^{-\lambda_n t}a_n r_n(x),
\end{equation}
where $\{\lambda_n\}_{n\ge 1}$ are the non-negative eigenvalues of $-\mathbb W$ in ascending order, and $\{r_n(x)\}_{n\ge 1}$ are the corresponding right eigenvectors. The coefficient $a_n$ is obtained by projecting the eigenvector $\ell_n(x)$ of ${\mathbb W}^\dagger$ onto the initial state, 
\begin{equation}\label{eq:defa2}
    a_n:=\int\dd x\,\ell_n(x) p(x,0).
\end{equation}
Because the dynamics Eq.~\eqref{eq:Langevin} satisfies the detailed balance property with respect to the Boltzmann distribution $\pi(x,\beta):=\ee^{-\beta V(x)}/Z(\beta)$ with $Z(\beta):=\int \dd x\, \ee^{-\beta V(x)}$, $p(x,t)$ converges to $\pi(x,\beta)$ at large times. 
In our study, the system is initially prepared in equilibrium at a temperature $T_i=\beta_i^{-1}$, so that the coefficients $a_n$ depend on $T_i$ in addition to $T$ and $V(x)$. 
While the lowest eigenvalue $\lambda_1=0$ is known ($\ell_1(x)=1$, $r_1(x)=\pi(x,\beta)$), the specifics of the spectrum are of course heavily potential-dependent~\cite{morsch1979one, risken1996fokker, walker2021anomalous, Biswas23a}, up to some robust properties. 
One of them is that at low bath temperature $T$, $\lambda_2\propto\ee^{-\beta \Delta V_{b}}$ in accordance with the Arrhenius law, as demonstrated by Kramers~\cite{arrhenius1889dissociationswarme,KRAMERS1940}, where $\Delta V_{b}$ is the potential barrier from the least stable well.
The corresponding timescale can be made arbitrarily large, leaving ample room to observe the Mpemba effect.
In the $t\gg\lambda_3^{-1}$ regime, the dynamics is dominated by the slowest eigenmode.
Thus, we can approximate
\begin{equation}
    p(x,t)\simeq \pi(x,\beta)+\ee^{-\lambda_2 t}a_2 r_2(x)+O(\ee^{-\lambda_3 t}),
\end{equation}
which we can use to determine how far $p(x,t)$ lies from $\pi(x,\beta)$. 
For instance, Refs.~\cite{lu2017nonequilibrium,kumar2020exponentially} quantify the distance from equilibrium using the Kullback--Leibler divergence $D(p\Vert\pi)(t)=\int\dd x\,p(x,t)\ln[p(x,t)/\pi(x,\beta)]$ (or an $L^1$ distance).
At large times, $D(t)\simeq \ee^{-2\lambda_2 t}a_{2}^{2}\int\dd x\, r^{2}_{2}(x)/\pi(x)$, and we see that the whole $T_i$-dependence is contained in $a_2$. 
The overall relaxation time scale is set by $\lambda_2^{-1}$, which is independent of initial conditions.
The initial condition dependence of the relaxation is thus encoded in $a_2$, and $\ell_{2}(x)$ reflects the contribution of state $x$ to the relaxation process. 
We fix the overall sign of $\ell_2(x)$ such that $\left.\partial a_2/\partial\beta_i\right|_{\beta_i=\beta}<0$.
Following the existing literature, we therefore diagnose the Mpemba effect through a nonmonotonic dependence of $a_2$ on the initial temperature $T_i$. We focus on a temperature $T_{\mathcal M}$ at which $\partial a_2/\partial\beta_i=0$; this Mpemba temperature marks the reversal of the slow-mode amplitude as the initial temperature is varied.
We now turn to an analytical study on the $T_i$-dependence of $a_2$.

Since we consider a regime where the bath temperature is low, up to an overall prefactor, we can approximate $\ell_2(x)$ by means of the following expression~\cite{matkowsky1977exit,bovier2004metastability, pre-us}:
\begin{equation}
    \begin{split}
        \ell_2(x)\simeq -&\frac{1+\ee^{-\beta \Delta V}}{2}\text{erf}\left[\sqrt{\frac{\beta |V''(x^*)|}{2}}(x-x^*)\right]\\
        +&\frac{1-\ee^{-\beta \Delta V}}{2},
    \end{split}
\end{equation}
where $x^*$ denotes the position of the local maximum of $V(x)$, and $\Delta V>0$ denotes the energy difference between the two wells. 
For our discussion, we introduce a boundary located at $x=-L_-(<0)$ to the left and one at $x=L_+$. In the low temperature limit, $\ell_2(x)$ is basically a step function $\theta(x^*-x)$. 
Besides, the derivative $\p a_2/\p \beta_i$ has a simple thermodynamic interpretation:
\begin{equation}\label{eq:derivativea2}
    \frac{\p a_2}{\p\beta_i}=\langle\ell_2(x)(U_i-V(x))\rangle_{\beta_i},
\end{equation}
where the brackets with $\beta_i$ refer to an equilibrium average with respect to $\pi(x,\beta_i)$, and $U_i=\int \dd x\,V(x)\pi(x,\beta_i)$ is the internal energy at temperature $T_i$. Our following analysis relies on the jump structure of $\ell_2(x)$, which is not unique to the double-well potentials, but also has analogues in single-well potentials~\cite{pre-us}.

We analyze a system with a bistable potential $V(x)$, where the two minima of $V(x)$ are located at $x_-$ and $x_+$ satisfying $V(x_-)>V(x_+)$ for $x_-<x_+$. In other words, the left well is the shallow one, and the right well is the deep one.
We establish that $a_2$ is a monotonic function of $\beta_i$ when $L_-=+\infty$, but that, for a large yet finite $L_-<L_+$, it possesses an extremum as a function of $\beta_i$. Due to Eq.~\eqref{eq:derivativea2}, the probabilistic interpretation of the Mpemba condition $\p a_2/\p\beta_i=0$ is that $V(x)$ and $\ell_2(x)$ are uncorrelated in the initial equilibrium. 
Physically, this means that the internal energies are equal in each domain. 
Indeed, denoting by $p_-:=\int_{-L_-}^{x^*}\dd x\, \pi(x,\beta_i)$ and $p_+:=\int_{x^*}^{L_+}\dd x\, \pi(x,\beta_i)$, we see that Eq.~\eqref{eq:derivativea2} amounts to 


\begin{equation}\label{eq:equalenergies}
   \underbrace{ \frac{1}{ p_-}\int_{-L_-}^{x^*}\dd x\,V(x)\pi(x,\beta_i)}_{U_-}=  \underbrace{\frac{1}{ p_+}\int_{x^*}^{L_+}\dd x\,V(x)\pi(x,\beta_i)}_{U_+},   
\end{equation}  
and it is now a matter of understanding the behavior of $\beta_i$ on each side. 
To be concrete, we use a generic confining polynomial potential that grows as $x^m$ for $|x|\gg 1$ with the form
\begin{equation}
    V(x)=v_{m}(x-x^*)^m+\sum_{k= 0}^{m-1} v_k (x-x^*)^k,   
\end{equation}
with even $m>2$ (with the $v_k$'s adjusted to comply with the asymmetric double well shape). 
For asymptotic estimates, we write this population as the contribution from the unbounded potential, minus the tail excluded by the left wall. After rescaling the first term, we obtain
\begin{equation}
\begin{split}
    Z(\beta_i) p_-&=\beta_i^{-1/m}\int_{-\infty}^{0}\dd x\,\ee^{-v_{m}x^m}\ee^{-\sum_{k=0}^{m-1}\beta_{i}^{1-\frac{k}{m}}v_k x^k}\\
    &~~-\int_{-\infty}^{-L_-}\dd x\,\ee^{-\beta_i V(x)},
\end{split}
\end{equation}
and  similar expressions are used not only for the right side but also for the numerators in Eq.~\eqref{eq:equalenergies}, which, after some tedious manipulations, lead to
\begin{equation}
    \label{eq:exprUminus}\begin{split}
        U_{-}=&\frac{1}{m\beta_{i}}+\frac{\sum_{k=0}^{m-1}(m-k)v_{k}\int_{-L_-}^{x^*}\dd x\,(x-x^*)^{k}\ee^{-\beta_{i} V(x)}}{m Z(\beta_i)p_{-}}\\
        &-\frac{(L_{-}+x^*)\ee^{-\beta_i V(-L_-)}}{m\beta_i Z(\beta_i)p_{-}},
    \end{split}
\end{equation}
and a similar expression holds for $U_+$. For both $L_-$ and $L_+$ large the condition Eq.~\eqref{eq:equalenergies} then leads to $\beta_i=\beta_{\mathcal M}$ with the Mpemba temperature given by
\begin{equation}\label{eq:transition2walls}
    \beta_{\mathcal M}\simeq\frac{C_1\ln (C_2 L_-)}{L_-^m}+\frac{1}{L_-^m}\ln\left(1-\frac{L_+}{L_-}\ee^{-\beta_{\mathcal M}(L_+^m-L_-^m)}\right),
\end{equation}
where $C_1$ and $C_2$ are constants. This inverse temperature indeed exists if $L_+$ is large enough with respect to $L_-$. Upon sending the right wall to infinity, we arrive at
\begin{equation}\label{eq:temperaturetransition}
\beta_{\mathcal M}\propto\frac{\ln L_-}{L_-^m}.
\end{equation}
The constant $C_2$ appearing in Eq.~\eqref{eq:transition2walls} arises from the asymmetry of the largest-order odd moment of $V$ on either side of $x^*$. 
Note that, for large $L_-$, the inner structure of the potential plays a minor role.
The details are provided in the End Matter.

Several comments are in order. First, we see from Eq.~\eqref{eq:temperaturetransition} that, in the absence of a right wall, sending the left boundary to infinity suppresses any possibility of the Mpemba effect since $\beta_{\mathcal M}\to 0$. This establishes that a boundary on the shallow side is a necessary condition. 
Somewhat paradoxically, the initial temperature regime we are studying in Eq.~\eqref{eq:temperaturetransition} is effectively a low-temperature regime on the left-hand side, since $\beta_{\mathcal M} V(-L_-)\sim\ln L_-\gg 1$, despite having $T_{\mathcal M}\to+\infty$. In addition, as shown in Eq.~\eqref{eq:transition2walls}, if a wall is included on the deep side, provided it is sufficiently far away, it plays no significant role. Nevertheless, bringing the right wall too close eliminates the Mpemba effect as inferred from Eq.~\eqref{eq:transition2walls}. 
Finally, note that the derivation of our results for large $L_-$ need not invoke the double-well structure of the potential, 
and our results directly extend to a single-well potential, as is detailed in \cite{pre-us}.

These analytical results establish a condition for the Mpemba effect in the low-bath-temperature regime. The finite-bath-temperature corrections decay exponentially with $\beta$, so we expect our results to hold well beyond the low-temperature limit. We have also used a boundary located far from the left well, which leads to a Mpemba temperature that increases with $L_-$. Whether the high $\beta$ and large $L_\pm$ results we have obtained are observed in practice for finite values of these parameters is now explored numerically.

\begin{figure}[b]
\begin{center}
\includegraphics[width=1\linewidth]{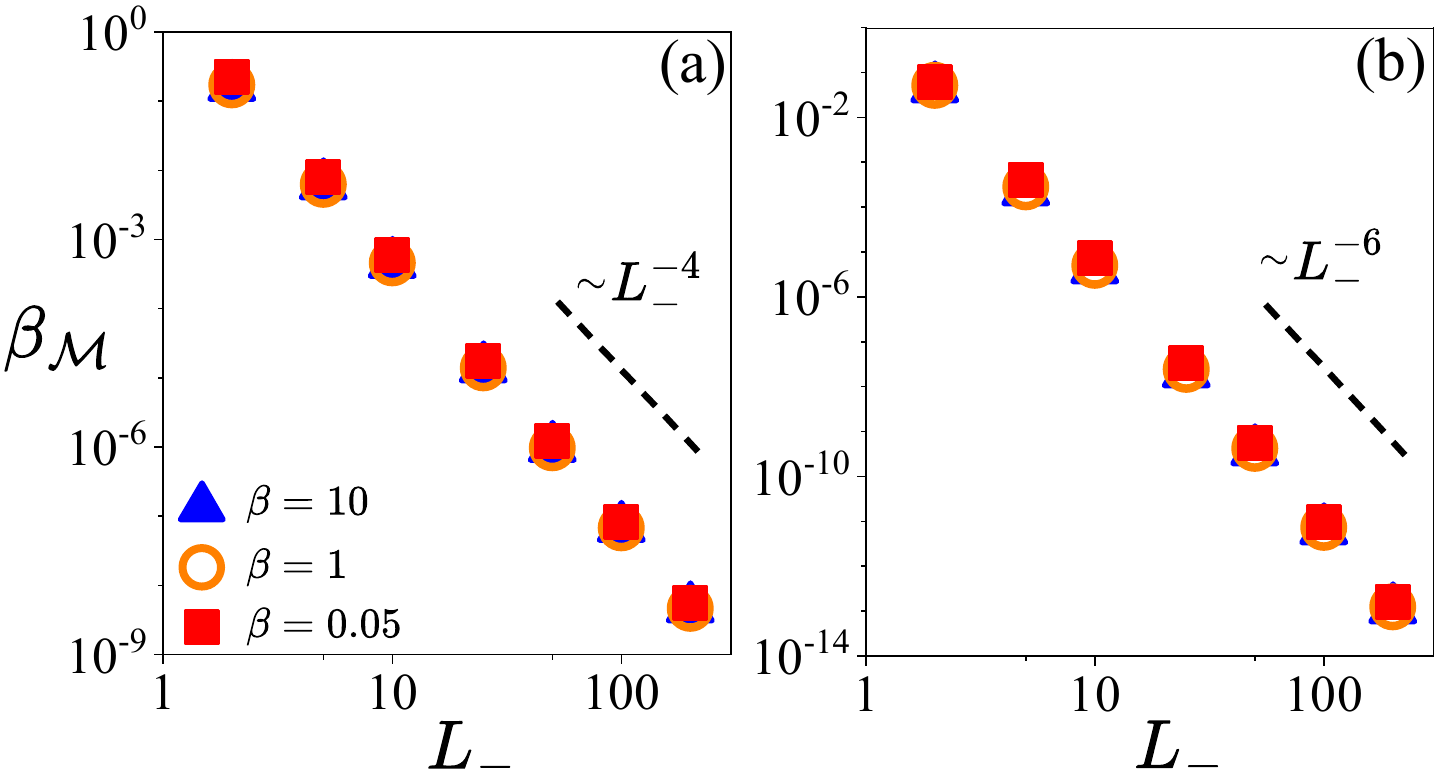}
\caption{The inverse Mpemba temperature $\beta_\mathcal{M}$ is shown as a function of $L_-$ for various bath temperatures in a quartic (a) and a sextic (b) potential. The analytical result obtained for large $\beta$ and $L_-$ is shown as a dashed line for guidance. 
No notable deviations are observed. We have used $L_+=2L_-$.}
\label{fig:betaivsbetaandL}
\end{center}
\end{figure}

We discretize the horizontal axis into $N=10^4$ lattice points, with the total length $[-L_-,L_+]$. 
We choose $L_+$ sufficiently large (in agreement with the existence of $\beta_{\mathcal M}$ in Eq.~\eqref{eq:transition2walls}) so that our results are independent of $ L_+$. The evolution operator $\mathbb W$ in Eq.~\eqref{eq:FP} is then represented as an $N\times N$ matrix. 
The latter is diagonalized using the {\tt scipy.linalg} package optimized to search for the first nonzero eigenvalue $\lambda_2$ and eigenvector $\ell_2(x)$. 
Then $a_2$ is determined according to Eq.~\eqref{eq:defa2} and the Mpemba temperature $\beta_{\mathcal M}$ is determined from the extremum of $a_2$, if it exists. For concreteness, we use a quartic potential of the form $V(x)=(x^2-1)^2-0.2 x$ and a sextic potential of the form $V(x)=(x^2-1)^2(x^2+1)-0.2 x$. The results for the Mpemba temperature $\beta_{\mathcal M}$ as a function of $L_-$ for various bath temperatures are shown in Fig.~\ref{fig:betaivsbetaandL}.

Although the formula in Eq.~\eqref{eq:temperaturetransition} was derived for large $L_-$ and large $\beta$, it is pragmatically valid down to $L_-$ of order unity and up to (pretty warm) bath temperatures comparable to four times the barrier height. Note that at very high bath temperature ($\beta=0.05$ in Fig.~\ref{fig:betaivsbetaandL}) the Mpemba temperature lies below the bath temperature, a phenomenon known as the inverse Mpemba effect~\cite{lu2017nonequilibrium}. 
We now propose a physical picture rationalizing our findings.

\begin{figure}[h]
\begin{center}
\includegraphics[width=0.45\textwidth]{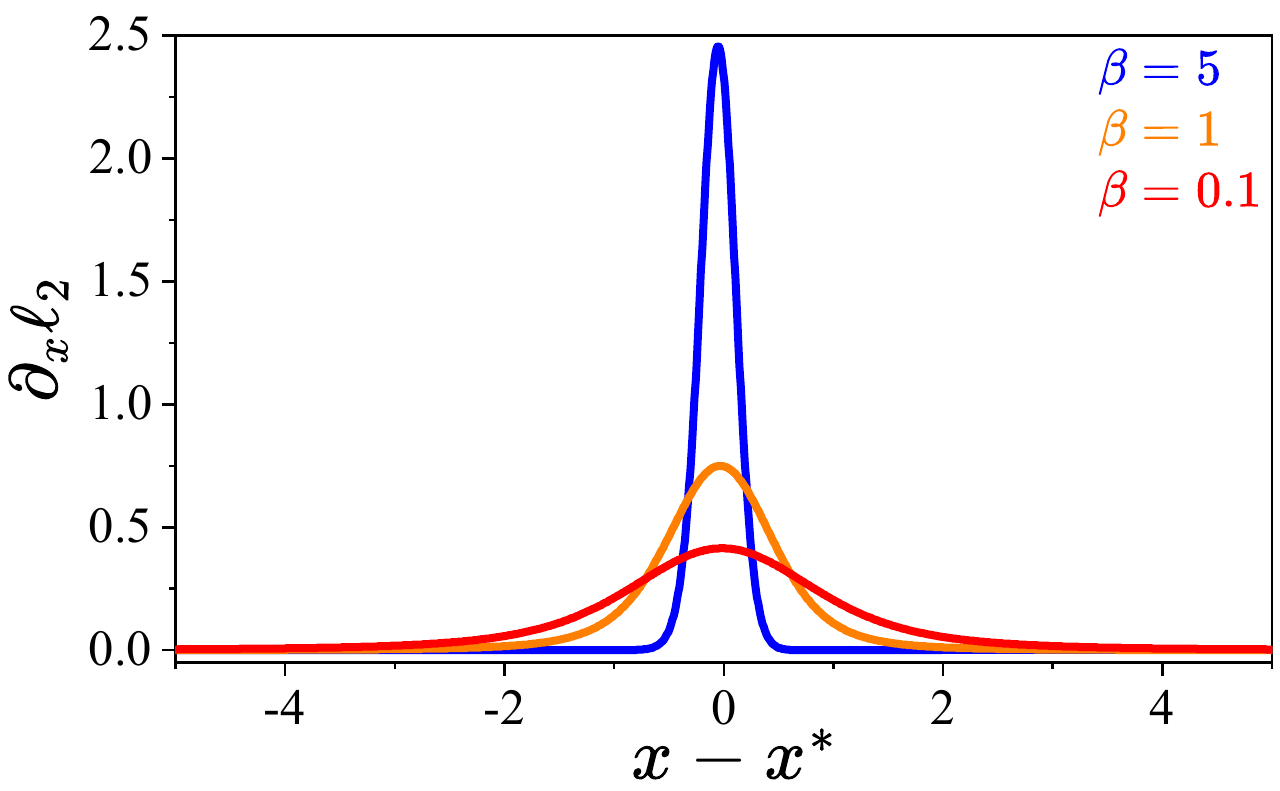}
\caption{The function $\p_x\ell_2$ exhibits a sharp peak at low bath temperature that is robust in the high temperature regime. Regardless of the temperature, it is flatly structureless far from the barrier.}
\label{fig:dxl2}
\end{center}
\end{figure}

The starting point of our interpretation is a rewriting of Eq.~\eqref{eq:derivativea2}
\begin{equation}
    \frac{\p a_2}{\p\beta_i}=\int\dd x\, \p_x\ell_2(x) \Delta U_i(x),
\end{equation}
where $\Delta U_i(x)=\int_{-\infty}^x\dd y\, (V(y)-U_i)\pi(y,\beta_i)$. 
By the Sturm-Liouville theory, the function $\ell_2(x)$ possesses exactly one zero and is monotonic (we choose $\p_x\ell_2(x)<0$). 
For all practical purposes (see Fig.~\ref{fig:dxl2} in the End Matter), $\p_x\ell_2(x)\simeq -\delta(x-x^*)$, so that the only ingredient that matters is the sign of $\Delta U_i(x^*)$ as a function of $\beta_i$ (the reasoning holds as long as the width of $\p_x\ell_2(x)$ is smaller than the distance to the wells or walls). 
We must understand the thermodynamics of $\Delta U_i(x^*)$ as a function of $T_i$. 

\begin{figure}[t]
\begin{center}

\includegraphics[width=\linewidth]{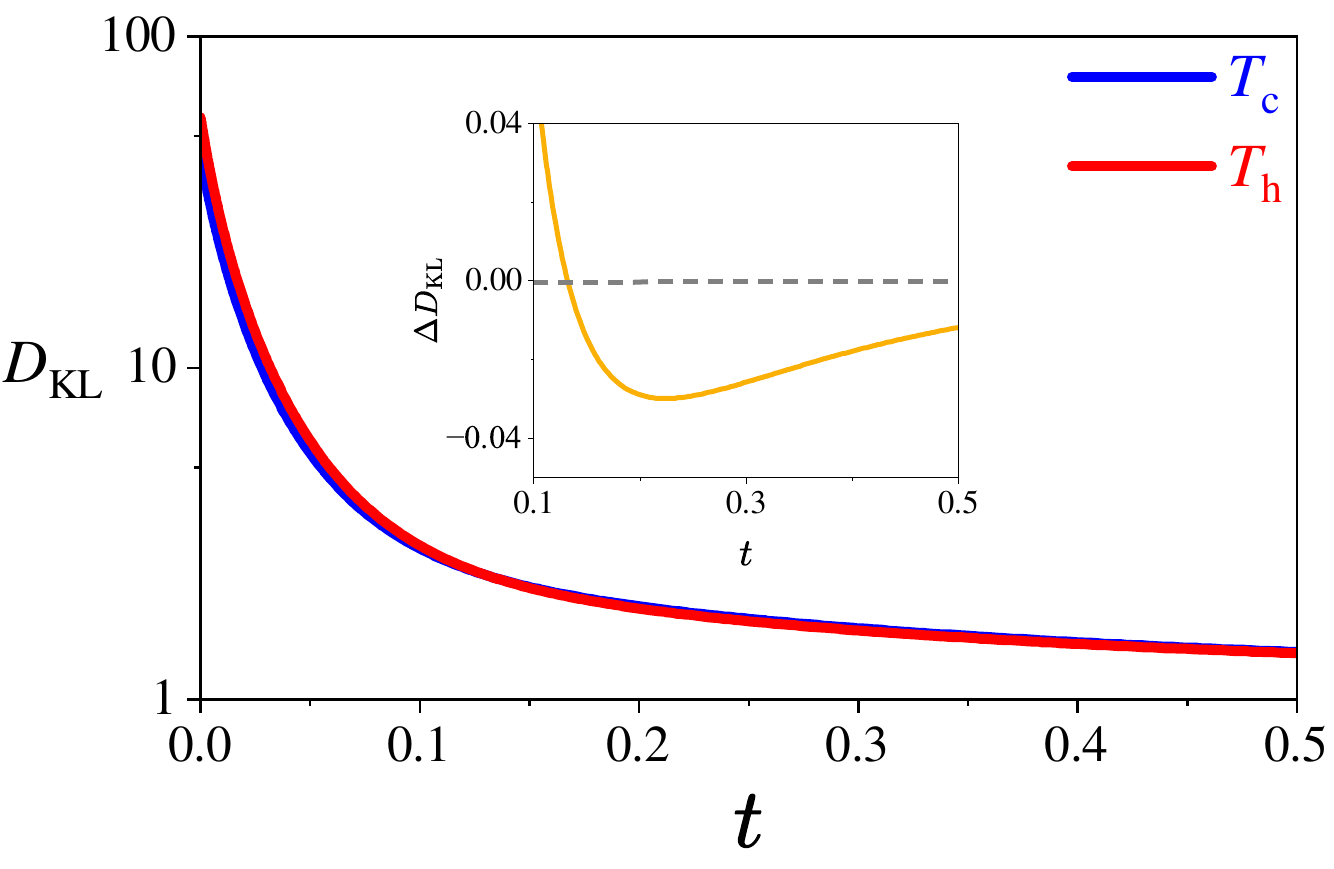}
\caption{Kullback--Leibler divergences for a hot initial temperature $T_h=T_{\mathcal M}+5$ and a cold initial temperature $T_c=T_{\mathcal M}-2$, chosen on opposite sides of $T_{\mathcal M}=24$, at bath temperature $T=0.1$. The corresponding $a_2$ coefficients differ by less than $0.1\%$. The inset shows $\Delta D_{\rm KL}=D_{\rm KL}(T_h)-D_{\rm KL}(T_c)$, which changes sign at $t_\times\simeq10^{-5}\lambda_2^{-1}$. The same potential as in Fig.~\ref{fig:betaivsbetaandL} is used.}
\label{fig:DKLdiff}
\end{center}
\end{figure}

We find it convenient to work directly at $L_+=+\infty$ initially. 
At low initial temperature $T_i$, most of the $\pi(x,\beta_i)$ population lies in the deepest minimum where $ U_i\simeq \min V$, so that $\Delta U_i(x^*)>0$ (for $y<x^*$, $V(y)>U_i$). 
This quantity will remain positive if we slightly increase the temperature, as the Boltzmann distribution starts filling the shallow well, even though $U_i$ increases. 
Then we consider a much higher initial temperature $T_i$ so that the Boltzmann distribution now feels the left wall. The distribution lowers and flattens, and begins to extend into the right wing. The contribution of $V$ (bounded to the left) is eventually overtaken by $U_i$ ({which grows without bound to infinity as $T_i\to\infty$}), making $\Delta U_i(x^*)$ negative. In terms of population, since $\Delta U_i(x^*)=T_i^2{\p p_-}/{\p T_i}$, we see that as $T_i$ increases, what causes the Mpemba effect is the transfer of the probability from the left side to the right one, where it can expand. Without such a wall to the left, the population would keep increasing on that side, hence killing the Mpemba effect. Now, let us insert a hard boundary to the right of the deep well. 
If the latter is placed too close to the minimum, the population reversal cannot occur, and $a_2$ is monotonic. 
Still, if, on the contrary, it is placed sufficiently far from the well, in such a way that, as $T_i$ increases, the Boltzmann distribution can sample the population in the vicinity of the left wall beforehand, then there must exist a Mpemba effect.
Beyond the low-temperature regime, the emergence of the Mpemba effect is still driven by the population transfer, which is in turn driven by the presence of a wall.
Moreover, the same population-transfer mechanism can operate in single-well and symmetric double-well potentials~\cite{pre-us}. Importantly, the resulting anomalous relaxation is not merely an asymptotic statement about the coefficient $a_2$. Figure~\ref{fig:DKLdiff} shows the full time evolution of the Kullback--Leibler divergence for two initial temperatures chosen on opposite sides of the maximum of $a_2$. Although their $a_2$ coefficients differ by less than $0.1\%$, the two relaxation curves exhibit a clear finite-time crossing at $t_\times\simeq10^{-5}\lambda_2^{-1}$, well before final equilibration.
The early crossing involves contributions from some faster relaxation modes ($t_{\times}\simeq 0.6\lambda_{3}^{-1}$), while the coefficient $a_2$ controls the late-time asymptotic ordering of the two curves~\cite{takadahayakawa}.

\begin{figure}[t]
\begin{center}

\includegraphics[width=1\linewidth]{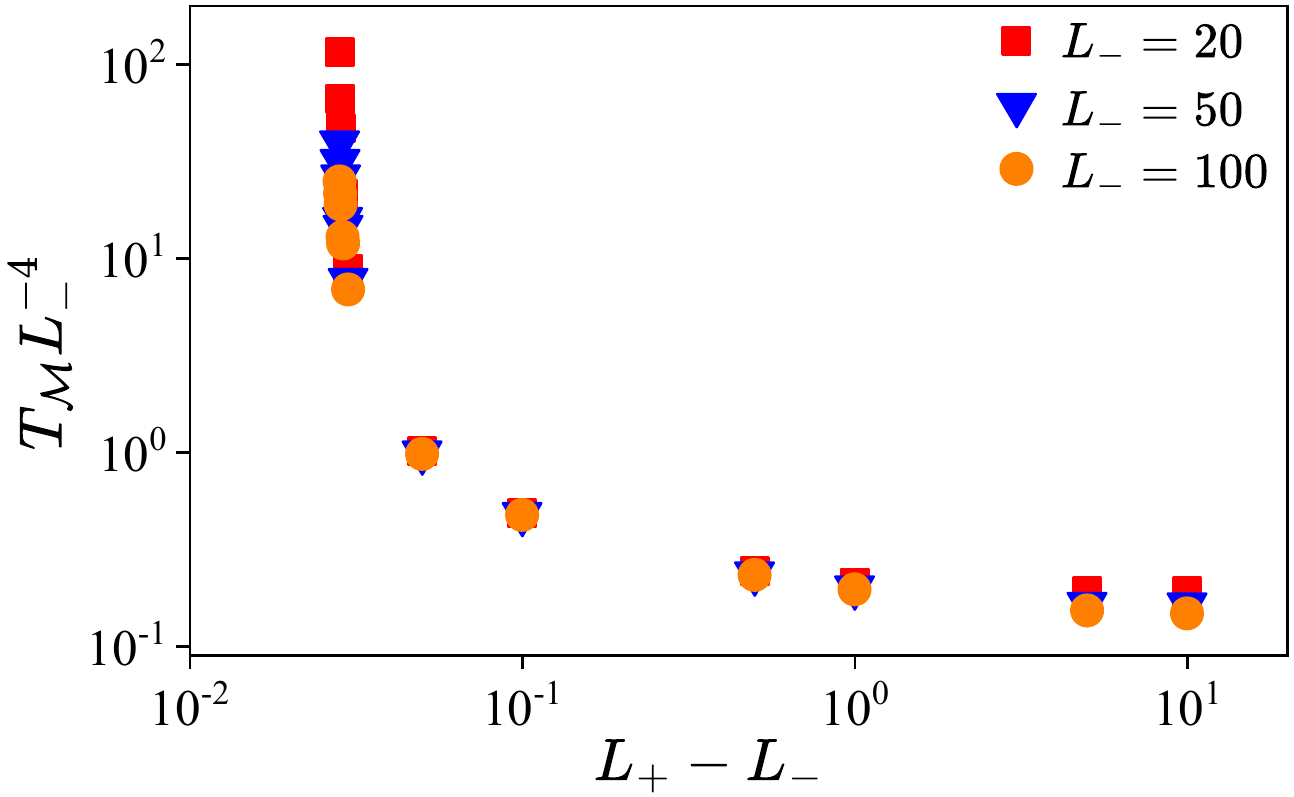}
\caption{The rescaled temperature $T_{\mathcal M}L_{-}^{-4}$ is shown as a function of $L_+-L_-$ for $L_-=20$ (red squares), $L_-=50$ (blue triangles) and $L_-=100$ (orange circles) when the bath is at $\beta=10$. The Mpemba temperature $T_{\mathcal M}$ diverges as the right wall gets close enough to the left wall, signaling the disappearance of the Mpemba effect. Our prediction Eq.~\eqref{eq:transition2walls} is consistent with the behavior away from $L_+\simeq L_-$.}
\label{fig:leftboundary}
\end{center}
\end{figure}

This analysis of the effect of a wall to the right, which is consistent with Eq.~\eqref{eq:transition2walls}, can also be explored numerically, as shown in Fig.~\ref{fig:leftboundary} for a quartic potential.
The rescaled Mpemba temperature $T_{\mathcal M}L_{-}^{-4}$ is shown as a function of the distance between the two walls, $L_+-L_-$.
As the right wall gets closer, $T_{\mathcal M}$ increases and eventually diverges, signaling the disappearance of the Mpemba effect.
For large $L_+-L_-$, the Mpemba temperature is independent of $L_+$ and is consistent with the prediction of Eq.~\eqref{eq:temperaturetransition}.
Although hard walls make the analytical mechanism particularly transparent, a literal boundary is not essential. A sufficiently asymmetric soft confinement can play the same role: when the shallow-side tail suppresses high-energy population more strongly than the opposite tail, increasing $T_i$ can reverse the population imbalance. The Supplemental Material~\cite{SM} treats the marginal case in which the two tails have the same power-law exponent but different prefactors. One can even construct potentials with two or more Mpemba temperatures by introducing additional high-energy confinement on the opposite side; an example is given in the End Matter (see also Refs.~\cite{Malhotra2024,pre-us}). These observations reinforce our central conclusion that the high-energy tails of the confining potential, rather than its local double-well structure alone, control the population inversion responsible for the effect.

In this work, we have used an asymmetric double-well potential to provide a long relaxation timescale on which the Mpemba effect can be cleanly resolved. For the polynomial potentials with hard boundaries considered analytically, we find that confinement on the shallow side is necessary: removing that boundary sends the Mpemba temperature to infinity. More generally, the same mechanism can be produced by sufficiently asymmetric soft tails, showing that the local double-well structure and metastability are not, by themselves, the ingredients responsible for the population inversion.
We have also shown that sufficiently strong confinement on the deep side can suppress the effect. Our results are rationalized by a population-transfer mechanism: as the initial temperature increases, the equilibrium distribution begins to probe the high-energy confinement, and the resulting redistribution between the two sides of the barrier can reverse the slow-mode amplitude.
Our findings now call for a quantitative systematic study~\cite{pre-us} of single- and double-well potentials, including cases in symmetric double-well potentials and in more than one space dimension without walls~\cite{takadahayakawa}, and situations where $a_2$ may not be the leading nonzero overlap, as in symmetric potentials. 
Our results suggest that the Mpemba effects observed in optical-trap experiments~\cite{kumar2020exponentially,tian2025experimental} may be sensitive to the high-energy confinement of the trap, rather than being determined by metastability alone. It would therefore be valuable to vary the far-tail shape of the optical potential independently of its local double-well structure and test how the anomalous relaxation changes.


\noindent\textbf{\textit{Note added after submission:}}
After we submitted this work, the authors of Ref.~\cite{sane2026mpemba} brought their study to our attention, prompting us to extend our results to cases where the left and right tails have the same power of energy but different prefactors, including the linear tails they considered (see SM~\cite{SM}).

\vspace{0.1in}
\noindent\textbf{\textit{Acknowledgements.}}
During this work, we became aware of a similar endeavor by John Bechhoefer and Siddharth Sane, with whom we had several useful exchanges that helped us clarify our message. 
We also thank Marija Vucelja and Apurba Biswas for several discussions.
HH thanks Satoshi Takada for fruitful discussions, and was supported by JSPS KAKENHI Grant No.~26K06960.
YL gratefully acknowledges the Yukawa Research Fellow, co-sponsored by the YITP and the Yukawa Memorial Foundation.
TVV was supported by JSPS KAKENHI Grant Nos.~JP23K13032, JP26K00019, JP26K00022, and JP26H02015. 
FvW acknowledges the financial support of the ANR grant THEMA No. 20-CE30-0031-01. 

\textit{Data availability.}---The data are not publicly available upon publication. The data are available from the authors upon reasonable request.

\bibliography{mpemba-gemini}
\onecolumngrid

\begin{center}
    {\large \bf End Matter}\\
\end{center}

\twocolumngrid
\raggedbottom

\noindent\textbf{\textit{Behavior of $\ell_2(x)$}:} 
The eigenvector $\ell_2$ of ${\mathbb W}^\dagger$ corresponding to $\lambda_2$ is known to have a single node. When $\lambda_2$ is quasi-degenerate with $0$, which occurs at sufficiently low bath temperature, it takes the form of a step function with a smooth error-function profile; hence, its derivative is a sharply peaked function near the location of the potential barrier. 
We have plotted in Fig.~\ref{fig:dxl2} for various values of $\beta$ the function $\p_x\ell_2(x)$ (normalized by the area under the peak $\int_{-\infty}^{+\infty}\dd y\,\p_y\ell_2(y)$). 
The potential we used is $V(x)=(x^2-1)^2-0.2 x$ (the wells are a distance unity from the center and the top of the barrier, and the potential barrier is also of order unity; hence $\beta=1$ refers to a temperature of the same order as the central barrier).

\noindent\textbf{\textit{Derivation of Eqs.~\eqref{eq:exprUminus} and \eqref{eq:transition2walls}}:} Due to Eq.~\eqref{eq:derivativea2}, the temperature $T_\mathcal{M}=\beta_\mathcal{M}^{-1}$, if it exists, is given by
\begin{equation}
    \langle \ell_{2}(x)V(x)\rangle_{\beta_\mathcal{M}}-\langle \ell_{2}(x)\rangle\langle V(x)\rangle_{\beta_\mathcal{M}}=0.
\end{equation}
At low bath temperature (with an exponentially small error in $\beta$), we use $\ell_2(x)\simeq\theta(x^*-x)$, hence
\begin{equation}
    U_{-}p_{-}-p_{-}(U_{-}p_{-}+U_{+}p_{+})=0 ,
\end{equation}
where $U_{\pm}$ and $p_{\pm}$ are defined by
\begin{equation}\label{eq:pandE}\begin{split}
    U_{\pm}:=&\pm\frac{1}{p_{\pm}}\int_{x^*}^{\pm\infty} \dd x\,V(x)\pi(x,\beta_{i}),\\
    p_{\pm}:=&\pm\int_{x^*}^{\pm\infty} \dd x\,\pi(x,\beta_{i}) ,
\end{split}\end{equation}
and by convention, $V(x)=+\infty$ for $x<-L_-$ and $x>L_+$. Then the Mpemba temperature solves
\begin{equation}
    U_{-}=U_{+}.
\end{equation}
The potential we consider can be written as
\begin{equation}
    V(x)=v_{m}(x-x^*)^{m}+\sum_{k=0}^{m-1}v_{k}(x-x^*)^{k} ,
\end{equation}
and it verifies
\begin{equation}
    \begin{split}
        (x-x^*)V'(x)=&mV(x)-\sum_{k=0}^{m-1}(m-k)v_{k}(x-x^*)^{k}.
    \end{split}
\end{equation}
By using the above identity and an integration by parts in Eq.~\eqref{eq:pandE}, we can get
\begin{equation}
    \begin{split}
        Z(\beta_i)p_{-}=&(L_{-}+x^*)\ee^{-\beta_{i} V(-L_-)}+\beta_{i}mU_{-}Z(\beta_i)p_{-}\\&-\beta_{i}\sum_{k=0}^{m-1}(m-k)v_{k}\int_{-L_-}^{x^*}\dd x\,(x-x^*)^{k}\ee^{-\beta_{i} V(x)},
    \end{split}
\end{equation}
and a similar expression holds for $p_+$. Thus we obtain
\begin{equation}
    \begin{split}
        U_{-}=&\frac{1}{m\beta_{i}}+\frac{\sum_{k=0}^{m-1}(m-k)v_{k}\int_{-L_-}^{x^*}\dd x\,(x-x^*)^{k}\ee^{-\beta_{i} V(x)}}{m Z(\beta_i)p_{-}}\\
        &-\frac{(L_{-}+x^*)\ee^{-\beta_i V(-L_-)}}{m\beta_i Z(\beta_i)p_{-}},\\
        U_{+}=&\frac{1}{m\beta_{i}}+\frac{\sum_{k=0}^{m-1}(m-k)v_{k}\int_{x^*}^{L_{+}}\dd x\,(x-x^*)^{k}\ee^{-\beta_{i} V(x)}}{mZ(\beta_i)p_{+}}\\
        &-\frac{(L_{+}-x^*)\ee^{-\beta_i V(L_+)}}{m\beta_i Z(\beta_i)p_{+}}.
    \end{split}
\end{equation}
For large enough $L_\pm$, the condition $U_{-}=U_{+}$ leads to
\begin{equation}
    \frac{L_-\ee^{-\beta_{{\mathcal M}} V(-L_-)}}{m\beta_{{\mathcal M}}p_{-}}-\frac{L_+\ee^{-\beta_{{\mathcal M}} V(L_+)}}{m\beta_{{\mathcal M}}p_{+}}
    =\sum_{k=0}^{m-1}\frac{m-k}{m}v_{k}\Delta\langle x^{k}\rangle,
\end{equation}
where $\Delta\langle x^{k}\rangle$ is defined as
\begin{equation}
    \Delta\langle x^{k}\rangle=\langle x^{k}\rangle_{+}-\langle x^{k}\rangle_{-}
\end{equation}
with
\begin{align}
   \langle x^{k}\rangle_{+}&=\frac{\int_{x^*}^{\infty}\dd x\,(x-x^*)^{k}\ee^{-\beta_{{\mathcal M}} V(x)}}{p_{+}} ,\\
   \langle x^{k}\rangle_{-}&=\frac{\int_{-L_-}^{x^*}\dd x\,(x-x^*)^{k}\ee^{-\beta_{\mathcal M} V(x)}}{p_{-}} .
\end{align}    
Our task now is to determine the order of $\Delta\langle x^{k}\rangle$. In the following, we discuss the large $L_-$ limit, and we anticipate that the regime of interest is the high-temperature one with small $\beta_{\mathcal M}$, yet with $\beta_{\mathcal M}V(-L_-)\gg 1$. In this limit, we have
\begin{equation}
    \begin{split}
    p_-&\simeq M_{0}-\frac{1}{m\beta_{\mathcal M}L_-^{m-1}}\ee^{-\beta_{\mathcal M} L_-^{m}},
    \end{split}
\end{equation}
where $M_{k}=\int_{0}^{\infty}\dd x\,x^{k}\ee^{-\beta_{{\mathcal M}} x^{m}}$. In addition we see that $M_{k}\sim \beta_{\mathcal M}^{-(k+1)/m}$. For the right and left sides, respectively, we have
\begin{equation}
    \begin{split}
        &\langle x^{k}\rangle_{+}\sim \frac{M_{k}}{M_{0}},\,\,\langle x^{k}\rangle_{-}\sim (-1)^k\frac{M_{k}}{M_{0}}.
    \end{split}
\end{equation}
We thus arrive at
\begin{equation}
\Delta\langle x^{k}\rangle\sim\begin{cases}
    \beta_{i}^{-k/m}&~\text{for odd}~k,\\
    0&~\text{for even}~k,
\end{cases}
\end{equation}
which suggests that only the highest odd term $k=k^*$ needs to be considered (this is $k^*=m-1$ if $v_{m-1} \neq 0$). Therefore, we have
\begin{equation}\label{eq:moment}
    \frac{L_-e^{-\beta_{\mathcal M} L_-^{m}}}{m\beta_{\mathcal M}\beta_{\mathcal M}^{-1/m}}-\frac{L_+\ee^{-\beta_{\mathcal M} L_+^{m}}}{m\beta_{\mathcal M}\beta_{\mathcal M}^{-1/m}}\sim \beta_{\mathcal M}^{-k^*/m},
\end{equation}
which establishes
\begin{equation}
    \begin{split}
    \beta_{\mathcal M}\simeq&(m-k^*)\frac{\ln (\text{const}\,L_-)}{L_-^m}\\
    &+\frac{1}{L_-^m}\ln\left(1-\frac{L_+}{L_-}\ee^{-\beta_{\mathcal M}(L_+^m-L_-^m)}\right),
    \end{split}
\end{equation}
irrespective of whether $k^*=m-1$ or a smaller integer.\\

\begin{figure}[H]
\begin{center}
\includegraphics[width=0.49\textwidth]{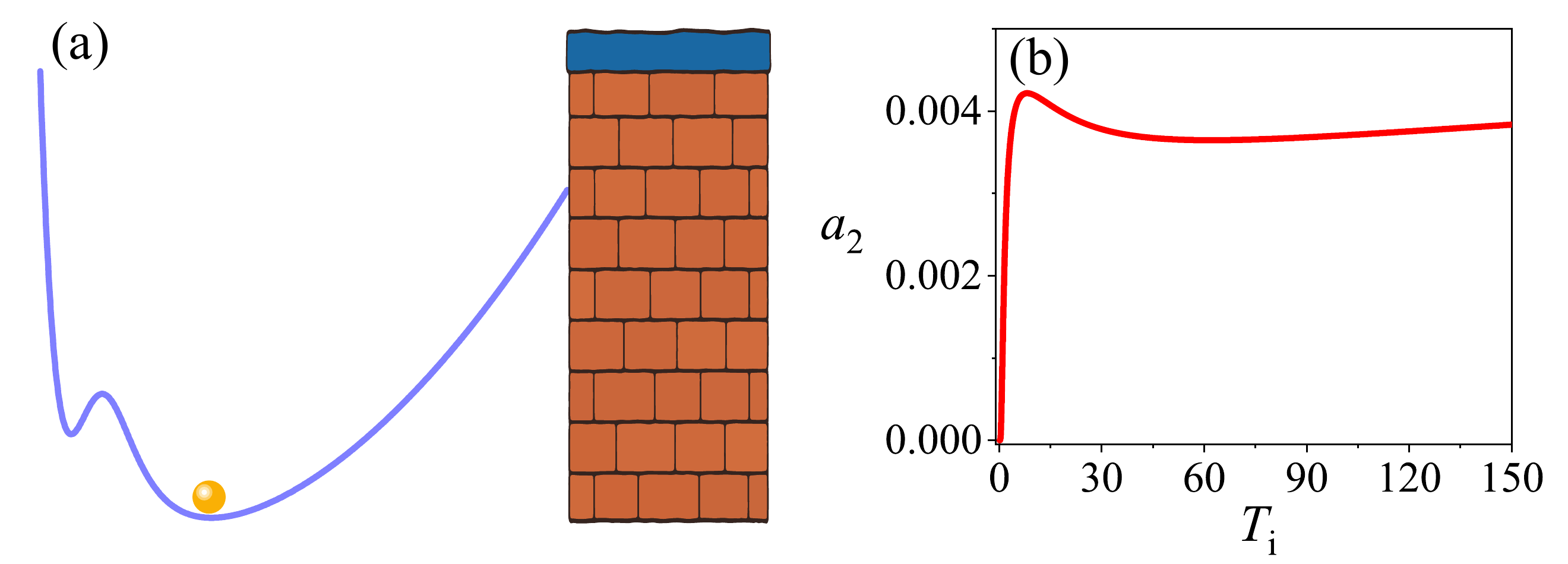}
\caption{For the choice $a=0.49$ in the definition of $V$ in Eq.~\eqref{eq:Vnona} and $L_+=40$, (a) the potential landscape is shown, and (b) the coefficient $a_2$ is shown as a function of the initial temperature $T_i$. This function exhibits two extrema.
\label{fig:nonanalytic}}
\end{center}
\end{figure}

\noindent\textbf{\textit{A mathematical curiosity}:} 
If the analyticity assumption is further relaxed, then, mostly as a mathematical game, one can artificially construct a potential that not only displays a Mpemba effect without the need for a hard wall (it only takes a steeper divergence), but also exhibits two (or more) Mpemba temperatures. 
Here, we adopt the choice 
\begin{equation}\label{eq:Vnona}
V(x)=\left\{
\begin{array}{l}
x^4-2x^2~~\text{for}~~x<0,\\
\dfrac{x^4}{1+a x^2}-2 x^2~~\text{for}~~x>0.\end{array}
\right.
\end{equation}
The left asymptotics $x^4$ is steeper than the right asymptotics $x^2$, which leads to the existence of a Mpemba effect. In addition, if a hard(er) wall is now inserted to the right, a secondary population inversion will occur, and a second Mpemba temperature exists. The numerical illustration of this phenomenon is provided in Fig.~\ref{fig:nonanalytic}.\\

Similar phenomena have been observed in Ref.~\cite{Degunther2022}. Here, we explain the underlying physics. Based on this mechanism, one can construct a potential with an arbitrary number of Mpemba temperatures even in single-well potentials (see the longer version~\cite{pre-us} for details).

\end{document}


\title{Supplemental Material for\texorpdfstring{\\}{: }``The Mpemba effect likes to hit a wall''}
\author{Yue Liu}
\affiliation{Center for Gravitational Physics and Quantum Information, Yukawa Institute for Theoretical Physics, Kyoto University, Kitashirakawa Oiwakecho, Sakyo-ku, Kyoto 606-8502, Japan}

\author{Tan Van Vu}
\affiliation{Center for Gravitational Physics and Quantum Information, Yukawa Institute for Theoretical Physics, Kyoto University, Kitashirakawa Oiwakecho, Sakyo-ku, Kyoto 606-8502, Japan}

\author{Raphaël Chétrite}

\affiliation{CNRS Laboratoire Ypatia des Sciences Mathématiques (LYSM), Piazzale Aldo Moro 5, 00185 Rome, Italy}

\author{Frédéric van Wijland}
\affiliation{Laboratoire Mati\`ere et Syst\`emes Complexes (MSC), Université Paris Cité  \& CNRS (UMR 7057), 75013 Paris, France}
\affiliation{Center for Gravitational Physics and Quantum Information, Yukawa Institute for Theoretical Physics, Kyoto University, Kitashirakawa Oiwakecho, Sakyo-ku, Kyoto 606-8502, Japan}

\author{Hisao Hayakawa}
\affiliation{Center for Gravitational Physics and Quantum Information, Yukawa Institute for Theoretical Physics, Kyoto University, Kitashirakawa Oiwakecho, Sakyo-ku, Kyoto 606-8502, Japan}
\maketitle

In the main text, we use hard walls to keep the analytical derivation transparent. A literal hard boundary, however, is not essential to the mechanism: it can be replaced by sufficiently steep confining tails. More generally, what matters is an asymmetry between the high-energy confinements on the two sides of the potential. Here we analyze a class of soft-wall potentials in which the left and right tails have the same power-law exponent but acquire different effective amplitudes through their matching to the asymmetric polynomial core. We show, for the nondegenerate case $q\neq m$, that the leading scaling of the Mpemba temperature is the same as for hard walls and that the correction caused by the spatial variation of the slow eigenfunction inside the tails is asymptotically negligible.\\

\section{Model and approximation scheme}

We consider a potential consisting of a polynomial core $V_0$ connected smoothly to two confining tails sharing the same power-law growth exponent $q\geq1$:
\begin{equation}
V(x)=
\begin{cases}
E_L+\dfrac{F_LL_-}{q}
\left[\left(\dfrac{-x}{L_-}\right)^q-1\right], & x<-L_-,\\[2mm]
V_0(x), & -L_-\leq x\leq L_+,\\[1mm]
E_R+\dfrac{F_RL_+}{q}
\left[\left(\dfrac{x}{L_+}\right)^q-1\right], & x>L_+.
\end{cases}
\label{eq:potential}
\end{equation}
The polynomial core is
\begin{equation}
V_0(x)=v_m(x-x^*)^m+\sum_{k=0}^{m-1}v_k(x-x^*)^k,
\qquad v_m>0,
\label{eq:core}
\end{equation}
where $m\geq4$ is even. The two tails have the same growth exponent $q\geq1$; $q=1$ corresponds to the linear tails considered in Ref.~\cite{sane2026mpemba}. We impose continuity of both $V$ and $V'$ at the junctions by choosing
\begin{equation}
\begin{aligned}
E_L&=V_0(-L_-), & F_L&=-V_0'(-L_-)>0,\\
E_R&=V_0(L_+), & F_R&=V_0'(L_+)>0.
\end{aligned}
\label{eq:matching}
\end{equation}
Equations~\eqref{eq:core} and \eqref{eq:matching} imply the exact identity
\begin{equation}
mE_L-F_L(L_-+x^*)=\sum_{k=0}^{m-1}(m-k)v_k(-L_--x^*)^k.
\label{eq:junction-matching-identity}
\end{equation}
and an analogous identity holds at the right junction. We assume that the junctions lie far from the central region of the core potential, $L_\pm\gg1$.
For $q=m$, the potential reduces to the no-wall case. The case without $C^1$ continuity is not considered here.

With soft confining tails, the initial equilibrium distribution extends beyond the junctions. The tail regions therefore contribute to the conditional energies, while the slow left eigenfunction $\ell_2$ need not remain exactly constant there. For hard walls, the low-temperature analysis in the main text justifies approximating $\ell_2$ by a step function with a jump near $x^*$. Here we retain this step-function contribution as the leading term and treat the variation of $\ell_2$ within the tails as a correction. Accordingly, we split $a_2$ and $\beta_{\mathcal M}$ into two contributions:
\begin{equation}
a_2=a_2^{(0)}+a_2^{(1)},
\qquad
\beta_{\mathcal M}=\beta_{\mathcal M}^{(0)}+\beta_{\mathcal M}^{(1)},
\label{eq:expansion}
\end{equation}
where $a_2^{(0)}$ denotes the step-function contribution and $a_2^{(1)}$ the correction generated by the variation of $\ell_2$ in the tails. The zeroth-order Mpemba inverse temperature $\beta_{\mathcal M}^{(0)}$ is defined by $\partial_{\beta_i}a_2^{(0)}=0$, while $\beta_{\mathcal M}^{(1)}$ denotes the shift of this root due to $a_2^{(1)}$. We verify below that this shift is asymptotically small. As in the main-text derivation, the zeroth-order stationarity condition leads to
\begin{equation}
    U_-=U_+.
\end{equation}
Here $U_\pm$ are the conditional mean energies on the two sides of $x^*$, including the corresponding tail contributions. We first determine the resulting zeroth-order Mpemba temperature and then show that the correction to $\partial a_2/\partial\beta_i$ is asymptotically negligible when the junctions are remote.\\

For fixed $q\neq m$ and a fixed asymmetric ratio $L_-/L_+\neq1$, the analysis below gives the inverse Mpemba temperature
\begin{equation}
\beta_{\mathcal M}^{(0)}\propto\frac{\ln L_-}{L_-^m},
\label{eq:leading-scaling}
\end{equation}
which has the same leading scaling as in the hard-wall case. We further show that $|\beta_{\mathcal M}^{(1)}|/\beta_{\mathcal M}^{(0)}\to0$ as $L_-\to\infty$, so the soft-tail correction does not modify this asymptotic result.

\section{Mpemba temperature in the step function approximation}

Let $Z=\int_{\mathbb R}\ee^{-\beta_i V}\dd x$, so that $Zp_-=\int_{-\infty}^{x^*}\ee^{-\beta_i V}\dd x$ and $U_-Zp_-=\int_{-\infty}^{x^*}V\ee^{-\beta_i V}\dd x$. Using $V'\ee^{-\beta_i V}=-\beta_i^{-1}\partial_x\ee^{-\beta_i V}$, an integration by parts on the full left half-line then gives
\begin{equation}
\begin{aligned}
(x-x^*)V_0'&=mV_0-\sum_{k=0}^{m-1}(m-k)v_k(x-x^*)^k,\\
\int_{-\infty}^{x^*}(x-x^*)V'\ee^{-\beta_i V}\dd x
&=-\frac{[(x-x^*)\ee^{-\beta_i V}]_{-\infty}^{x^*}}{\beta_i}
+\frac1{\beta_i}\int_{-\infty}^{x^*}\ee^{-\beta_i V}\dd x
=\frac{Zp_-}{\beta_i}.
\end{aligned}
\label{eq:integration-identities}
\end{equation}
The boundary terms vanish at $x^*$ and $-\infty$.\\

At the junction with the left tail we have $xV'=q(V-E_L)+F_LL_-$, hence
\begin{equation}
\begin{aligned}
(x-x^*)V'
&=xV'-x^*V'\\
&=q(V-E_L)+F_LL_--x^*V'\\
&=mV-\bigl[(m-q)(V-E_L)+(mE_L-F_LL_-)+x^*V'\bigr],
\end{aligned}
\label{eq:left-tail-identity}
\end{equation}
where we have used $q(V-E_L)=mV-(m-q)(V-E_L)-mE_L$ in the last line. Using the first lines of Eqs.~\eqref{eq:integration-identities} and \eqref{eq:left-tail-identity} leads to the  generalization of Eq.~(21) in the main text:
\begin{equation}
\begin{aligned}
Zp_-={}&m\beta_i U_-Zp_-\\
&-\beta_i\sum_{k=0}^{m-1}(m-k)v_k
\int_{-L_-}^{x^*}(x-x^*)^k\ee^{-\beta_i V_0}\dd x\\
&-\beta_i\int_{-\infty}^{-L_-}
\bigl[(m-q)(V-E_L)+(mE_L-F_LL_-)+x^*V'\bigr]\ee^{-\beta_i V}\dd x.
\end{aligned}
\label{eq:exact-left-energy-relation}
\end{equation}

For junctions to potentials with tails characterized by a fixed $q\neq m$ and fixed $L_-/L_+\neq1$, with $L_\pm\to\infty$, we {\it a priori} search for a correction $\beta_i\to0$ with $\beta_i E_{L,R}\gg1$. In this regime, the leading tail integrals are (see Sec.~\ref{sec:endpoint-details})
\begin{equation}
\int_{-\infty}^{-L_-}\ee^{-\beta_i V}\dd x
\simeq\frac{\ee^{-\beta_i E_L}}{\beta_i F_L},
\qquad
\int_{-\infty}^{-L_-}(V-E_L)\ee^{-\beta_i V}\dd x
\simeq\frac{\ee^{-\beta_i E_L}}{\beta_i^2F_L},
\label{eq:tail-estimates}
\end{equation}
up to $O(1/\beta_i E_L)$ corrections. The last term in the third line of Eq.~\eqref{eq:exact-left-energy-relation}  integrates exactly into $\int_{-\infty}^{-L_-}x^*V'\ee^{-\beta_i V}\dd x=-x^*\ee^{-\beta_i E_L}/\beta_i$. Thus the constant and derivative terms in the tail integral of Eq.~\eqref{eq:exact-left-energy-relation} add up into  $[mE_L-F_L(L_-+x^*)]\ee^{-\beta_i E_L}/(\beta_i F_L)$ to leading order.

The corresponding endpoint estimate for the full polynomial continuation is
(see Sec.~\ref{sec:endpoint-details})
\begin{equation}
\int_{-\infty}^{-L_-}(x-x^*)^k\ee^{-\beta_i V_0}\dd x
\simeq(-L_--x^*)^k\frac{\ee^{-\beta_i E_L}}{\beta_i F_L}.
\label{eq:junction-cancellation}
\end{equation}
Using the matching identity~\eqref{eq:junction-matching-identity}, the junction terms in Eq.~\eqref{eq:exact-left-energy-relation} supply these missing moments, extending the core integrals to $(-\infty,x^*]$. The excess-energy tail term remains, giving the following leading relation (see Sec.~\ref{sec:endpoint-details} for the term-by-term calculation):
\begin{equation}
\begin{aligned}
Zp_-\simeq{}&m\beta_i U_-Zp_-\\
&-\beta_i\sum_{k=0}^{m-1}(m-k)v_k
\int_{-\infty}^{x^*}(x-x^*)^k\ee^{-\beta_i V_0}\dd x
-\frac{m-q}{\beta_i F_L}\ee^{-\beta_i E_L}.
\end{aligned}
\label{eq:left-energy-relation}
\end{equation}
Solving for $U_-$ yields
\begin{equation}
\begin{aligned}
U_-\simeq{}&\frac{1}{m\beta_i}
+\frac{1}{mZp_-}\sum_{k=0}^{m-1}(m-k)v_k
\int_{-\infty}^{x^*}(x-x^*)^k\ee^{-\beta_i V_0}\dd x\\
&+\frac{m-q}{m\beta_i^2F_LZp_-}\ee^{-\beta_i E_L}.
\end{aligned}
\label{eq:left-conditional-energy}
\end{equation}
The corresponding expression for $U_+$ is obtained by replacing the left half-line by the right half-line and $(E_L,F_L,p_-)$ by $(E_R,F_R,p_+)$.

To exploit the condition $U_-=U_+$, we use the moments estimates
\begin{equation}
M_k=\int_0^\infty y^k\ee^{-\beta_i v_my^m}\dd y
\propto\beta_i^{-(k+1)/m},
\qquad Zp_\pm\simeq M_0,
\label{eq:moments}
\end{equation}
where $y$ is distance from $x^*$. The left and right polynomial moments are $(-1)^kM_k$ and $M_k$ to leading order. The contribution $1/(m\beta_i)$ cancels out in the two energies and multiplying by $mM_0$ leads us to
\begin{equation}
0\simeq{}\sum_{k=0}^{m-1}(m-k)v_k[(-1)^k-1]M_k+\frac{m-q}{\beta_i^2}
\left[\frac{\ee^{-\beta_i E_L}}{F_L}
-\frac{\ee^{-\beta_i E_R}}{F_R}\right].
\label{eq:energy-balance}
\end{equation}
The leading even moments cancel each other out. Retaining the highest odd index $k^*$ with $v_{k^*}\neq0$, multiplying by $\beta_i$, and setting $\beta_i=\beta_{\mathcal M}^{(0)}$ yields
\begin{equation}
\frac{m-q}{\beta_{\mathcal M}^{(0)}}
\left[\frac{\ee^{-\beta_{\mathcal M}^{(0)}E_R}}{F_R}
-\frac{\ee^{-\beta_{\mathcal M}^{(0)}E_L}}{F_L}\right]
\simeq-2(m-k^*)v_{k^*}\beta_{\mathcal M}^{(0)}M_{k^*}.
\label{eq:leading-temperature-equation}
\end{equation}
Equation~\eqref{eq:leading-temperature-equation} replaces Eq.~(30) in the main text in the case of soft tails. The moment $M_{k^*}$ is now evaluated at $\beta_{\mathcal M}^{(0)}$. The right-hand side scales as $(\beta_{\mathcal M}^{(0)})^{(m-k^*-1)/m}$ and reduces to $-2v_{m-1}/(mv_m)$ when $k^*=m-1$.

For $v_{k^*}<0$, $q>m$, and a fixed ratio $L_-/L_+<1$, compare the two tail terms in Eq.~\eqref{eq:leading-temperature-equation}:
\[
\frac{e^{-\beta_{\mathcal M}^{(0)}E_R}/F_R}
{e^{-\beta_{\mathcal M}^{(0)}E_L}/F_L}
=\frac{F_L}{F_R}
\exp\!\left[-\beta_{\mathcal M}^{(0)}E_L
\left(\frac{E_R}{E_L}-1\right)\right].
\]
At remote junctions,
\[
\frac{F_L}{F_R}\simeq\left(\frac{L_-}{L_+}\right)^{m-1},
\qquad
\frac{E_R}{E_L}\simeq\left(\frac{L_+}{L_-}\right)^m>1.
\]
In the regime $\beta_{\mathcal M}^{(0)}E_L\to\infty$, the ratio of the two tail terms therefore tends to zero, so the left exponential dominates.
Equation~\eqref{eq:leading-temperature-equation} then gives
\begin{equation}
(\beta_{\mathcal M}^{(0)})^{(2m-k^*-1)/m}e^{\beta_{\mathcal M}^{(0)}E_L}
\simeq\frac{\mathrm{const}}{F_L},
\label{eq:single-tail-temperature-equation}
\end{equation}
with a positive constant. Taking logarithms and using $E_L\simeq v_mL_-^m$,
$F_L\simeq mv_mL_-^{m-1}$ gives
\begin{equation}
\beta_{\mathcal M}^{(0)}E_L+
\frac{2m-k^*-1}{m}\ln(\beta_{\mathcal M}^{(0)}E_L)
=(m-k^*)\ln L_-+O(1).
\label{eq:log-temperature-equation}
\end{equation}
Keeping the leading logarithm, as in Eq.~(31) of the main text, we obtain
\begin{equation}
\beta_{\mathcal M}^{(0)}\propto\frac{\ln L_-}{L_-^m}.
\label{eq:final-temperature}
\end{equation}
For $q<m$ and $L_+<L_-$, the right exponential dominates. Thus, within the present $q\neq m$ asymptotic regime, the soft-tail problem has the same leading scaling as the hard-wall problem.

\section{Correction to the Mpemba temperature}

We now estimate the correction caused by the variation of $\ell_2$ within the soft tails. Let $B_L=\ell_2(-L_-)$ and $B_R=\ell_2(L_+)$, with $B_L>0>B_R$. The step-function contribution is $a_2^{(0)}=B_Lp_-+B_Rp_+$, whereas the tail-dependent correction is $a_2^{(1)}=Z^{-1}\sum_{j=L,R}\int_{\mathrm{tail}\,j}(\ell_2-B_j)\ee^{-\beta_i V}\dd x$. Its derivative with respect to the initial inverse temperature is exactly
\begin{equation}
\frac{\partial a_2^{(1)}}{\partial\beta_i}
=\frac1Z\sum_{j=L,R}\int_{\mathrm{tail}\,j}
(U_i-V)(\ell_2-B_j)\ee^{-\beta_i V}\dd x,
\qquad U_i=\langle V\rangle_{\beta_i}.
\label{eq:tail-projection-derivative}
\end{equation}
The condition becomes $\partial_{\beta_i}a_2\simeq (B_L-B_R)p_-p_+(U_+-U_-)+\partial_{\beta_i}a_2^{(1)}=0$.\\

For such convex tails as we are considering, comparison with a constant restoring force $F_j$ gives that at low enough bath temperature we have $F_j^2>4\lambda_2/\beta$ (see the calculational details in Sec.~\ref{sec:bound-details}),
\begin{equation}
|\ell_2-B_j|
\le |B_j|\frac{2\lambda_2(V-E_j)}{F_j^2}
\exp\!\left[\frac{2\lambda_2(V-E_j)}{F_j^2}\right].
\label{eq:tail-mode-bound}
\end{equation}
We change variables in Eq.~\eqref{eq:tail-projection-derivative} by setting $u=\beta_i(V-E_j)$, and we use $|U_i|\le1/\beta_i$ for sufficiently remote junctions in the
temperature regime of interest along with $2\lambda_2/(\beta_i F_j^2)\le1/2$, which brings us to
\begin{equation}
\left|\frac{\partial a_2^{(1)}}{\partial\beta_i}\right|
\le\frac{2\lambda_2}{Z}\sum_{j=L,R}
\frac{|B_j|\ee^{-\beta_i E_j}}{\beta_i^2F_j^3}
(E_j+\beta_i^{-1})
\int_0^\infty u(1+u)\ee^{-u/2}\dd u.
\label{eq:dimensionless-tail-bound}
\end{equation}
where the integral on the right-hand side equals $20$. Using $|B_j|\le B_L-B_R$ now leads to
\begin{equation}
\left|\frac{\partial a_2^{(1)}}{\partial\beta_i}\right|
\le\frac{40(B_L-B_R)\lambda_2}{Z\beta_i}
\sum_{j=L,R}\frac{\ee^{-\beta_i E_j}}{\beta_i F_j}
\frac{E_j+\beta_i^{-1}}{F_j^2}.
\label{eq:scaled-tail-bound}
\end{equation}
On either branch derived above, Eqs.~\eqref{eq:leading-temperature-equation} and \eqref{eq:final-temperature} imply
\begin{equation}
\frac{E_j+(\beta_{\mathcal M}^{(0)})^{-1}}{F_j^2}=O(L_-^{2-m}),
\qquad
\sum_{j=L,R}\frac{\ee^{-\beta_{\mathcal M}^{(0)}E_j}}
{\beta_{\mathcal M}^{(0)}F_j}
=O\!\left((\beta_{\mathcal M}^{(0)})^{(m-k^*-1)/m}\right).
\label{eq:tail-orders-at-root}
\end{equation}
Using also that $Z$ is of order $(\beta_{\mathcal M}^{(0)})^{-1/m}$ in Eq.~\eqref{eq:scaled-tail-bound},
we obtain the following bound, where $C$ collects the asymptotic constants
and is independent of $\beta_i$ and of the distances to the  junctions:
\begin{equation}
\left|\left.\frac{\partial a_2^{(1)}}{\partial\beta_i}
\right|_{\beta_i=\beta_{\mathcal M}^{(0)}}\right|
\le 40C(B_L-B_R)\lambda_2
(\beta_{\mathcal M}^{(0)})^{-k^*/m}L_-^{2-m}.
\label{eq:relative-tail-correction}
\end{equation}
In the step function approximation of $\ell_2$, $\partial_{\beta_i}a_2^{(0)}=(B_L-B_R)p_-p_+(U_+-U_-)$. Since $p_\pm\simeq1/2$ and $U_{+}-U_{-}\simeq M_{k^*}/M_0$ is of order $(\beta_{\mathcal M}^{(0)})^{-k^*/m}$, the core asymmetry and the balancing tail term each contribute on the scale $(B_L-B_R)(\beta_{\mathcal M}^{(0)})^{-k^*/m}$ to this derivative. Dividing Eq.~\eqref{eq:relative-tail-correction} by this scale gives a relative correction $O(\lambda_2L_-^{2-m})$. The same estimate holds near the solution $\beta_{\mathcal M}$ thanks to Eq.~\eqref{eq:scaled-tail-bound}.\\

For the left branch, we apply the same leading balance of Eq.~\eqref{eq:leading-temperature-equation} to the zeroth-order and to the stationarity condition of $a_2$ at $\beta_{\mathcal M}$:
\begin{equation}
\begin{aligned}
(\beta_{\mathcal M}^{(0)})^{(2m-k^*-1)/m}
 \ee^{E_L\beta_{\mathcal M}^{(0)}}
&=\frac{\mathrm{const}}{F_L},\\
\beta_{\mathcal M}^{(2m-k^*-1)/m}\ee^{E_L\beta_{\mathcal M}}
&=\frac{\mathrm{const}}{F_L}
\left[1+O(\lambda_2L_-^{2-m})\right],
\end{aligned}
\label{eq:perturbed-temperature-balance}
\end{equation}
with the same positive constant independent of $\beta_i$. Dividing the second equation by the first and using $\beta_{\mathcal M}=\beta_{\mathcal M}^{(0)}+\beta_{\mathcal M}^{(1)}$ gives, to leading order,
\begin{equation}
\left(1+\frac{\beta_{\mathcal M}^{(1)}}{\beta_{\mathcal M}^{(0)}}\right)^{(2m-k^*-1)/m}
\ee^{E_L\beta_{\mathcal M}^{(1)}}
=1+O(\lambda_2L_-^{2-m}).
\label{eq:temperature-ratio}
\end{equation}
Taking logarithms,
\begin{equation}
E_L\beta_{\mathcal M}^{(1)}
+\frac{2m-k^*-1}{m}
\ln\!\left(1+\frac{\beta_{\mathcal M}^{(1)}}{\beta_{\mathcal M}^{(0)}}\right)
=O(\lambda_2L_-^{2-m}),
\label{eq:log-temperature-shift}
\end{equation}
and using that the two terms on the left have the same sign, we thus have $E_L|\beta_{\mathcal M}^{(1)}|\lesssim\lambda_2L_-^{2-m}$. Since $\beta_{\mathcal M}^{(0)}E_L$ is of order $\ln L_-$, we obtain
\begin{equation}
\frac{|\beta_{\mathcal M}^{(1)}|}{\beta_{\mathcal M}^{(0)}}
\lesssim\frac{\lambda_2L_-^{2-m}}{\ln L_-}\longrightarrow0.
\label{eq:relative-temperature-shift}
\end{equation}
Equation~\eqref{eq:relative-temperature-shift} therefore shows that the soft-tail-induced shift of the Mpemba temperature is asymptotically negligible, so the leading hard-wall scaling is unchanged.

\section{Calculational details of the asymptotic estimates}
\label{sec:details}
This section provides explicit derivations of the asymptotic estimates used above to show that the soft power-law tails do not alter the leading behavior of $\beta_{\mathcal M}$.
\subsection{Endpoint estimates and matching cancellation}
\label{sec:endpoint-details}

We derive Eq.~\eqref{eq:tail-estimates} directly from the tail potential in Eq.~\eqref{eq:potential}. Set $u=\beta_i(V-E_L)$ and $\varepsilon=(\beta_i F_LL_-)^{-1}$. The exact inverse relation is $-x=L_-(1+q\varepsilon u)^{1/q}$, with Jacobian $|\dd x|=(\beta_i F_L)^{-1}(1+q\varepsilon u)^{1/q-1}\dd u$. Hence
\begin{equation}
\begin{aligned}
\int_{-\infty}^{-L_-}\ee^{-\beta_i V}\dd x
&=\frac{\ee^{-\beta_i E_L}}{\beta_i F_L}
\int_0^\infty \ee^{-u}(1+q\varepsilon u)^{1/q-1}\dd u,\\
\int_{-\infty}^{-L_-}(V-E_L)\ee^{-\beta_i V}\dd x
&=\frac{\ee^{-\beta_i E_L}}{\beta_i^2 F_L}
\int_0^\infty ue^{-u}(1+q\varepsilon u)^{1/q-1}\dd u.
\end{aligned}
\label{eq:app-exact-tail-integrals}
\end{equation}
For fixed $q\ge1$, Taylor's theorem gives
\begin{equation}
\begin{aligned}
(1+q\varepsilon u)^{1/q-1}
&=1-(q-1)\varepsilon u+R(u),\\
|R(u)|&\le\frac{(q-1)(2q-1)}2\varepsilon^2u^2,
\qquad u\ge0.
\end{aligned}
\label{eq:app-tail-taylor-bound}
\end{equation}
The bound holds for all $u\ge0$: the second derivative of $(1+q\varepsilon u)^{1/q-1}$ with respect to $u$ is nonnegative and bounded by its value $(q-1)(2q-1)\varepsilon^2$ at $u=0$. Integrating the remainder against $\ee^{-u}$ or $u\ee^{-u}$ therefore gives an $O(\varepsilon^2)$ error. Using $\int_0^\infty u^r e^{-u}\dd u=r!$ for nonnegative integers $r$, we obtain
\begin{equation}
\begin{aligned}
\int_{-\infty}^{-L_-}\ee^{-\beta_i V}\dd x
&=\frac{\ee^{-\beta_i E_L}}{\beta_i F_L}
\bigl[1-(q-1)\varepsilon+O(\varepsilon^2)\bigr],\\
\int_{-\infty}^{-L_-}(V-E_L)\ee^{-\beta_i V}\dd x
&=\frac{\ee^{-\beta_i E_L}}{\beta_i^2 F_L}
\bigl[1-2(q-1)\varepsilon+O(\varepsilon^2)\bigr].
\end{aligned}
\label{eq:app-tail-integral-expansion}
\end{equation}
At the junctions, $F_LL_-\simeq mE_L$, so $\varepsilon\simeq(m\beta_i E_L)^{-1}\ll1$. Keeping the leading terms gives Eq.~\eqref{eq:tail-estimates}, with relative errors $O((\beta_i E_L)^{-1})$. For $q=1$ the leading expressions are exact. The right-tail estimates follow identically.

We now explain how to compute the integrals in Eq.~\eqref{eq:junction-cancellation}. With one integration by parts, for $0\le k\le m-1$, we obtain
\begin{equation}
\begin{aligned}
\int_{-\infty}^{-L_-}(x-x^*)^k \ee^{-\beta_i V_0}\dd x
={}&\frac{(-L_--x^*)^k \ee^{-\beta_i E_L}}{\beta_i F_L}\\
&+\frac1{\beta_i}\int_{-\infty}^{-L_-}
\left(\frac{(x-x^*)^k}{V_0'}\right)'\ee^{-\beta_i V_0}\dd x.
\end{aligned}
\label{eq:app-polynomial-integration-by-parts}
\end{equation}
The boundary term uses $V_0(-L_-)=E_L$ and $V_0'(-L_-)=-F_L$; the term at $-\infty$ vanishes. The leading polynomial term gives
\begin{equation}
\frac{\displaystyle\left(\frac{(x-x^*)^k}{V_0'}\right)'}{(x-x^*)^k}
=\frac{k}{(x-x^*)V_0'}-\frac{V_0''}{(V_0')^2}
=O(|x|^{-m}).
\label{eq:app-polynomial-remainder}
\end{equation}
Equation~\eqref{eq:app-polynomial-remainder} also holds within the tail. Since $|x|\ge L_-$, the remainder in Eq.~\eqref{eq:app-polynomial-integration-by-parts} satisfies
\begin{equation}
\begin{aligned}
\left|\frac1{\beta_i}\int_{-\infty}^{-L_-}
\left(\frac{(x-x^*)^k}{V_0'}\right)'\ee^{-\beta_i V_0}\dd x\right|
&\le\frac1{\beta_i}\int_{-\infty}^{-L_-}
\left|\left(\frac{(x-x^*)^k}{V_0'}\right)'\right|\ee^{-\beta_i V_0}\dd x\\
&\le\frac{C}{\beta_iL_-^m}\int_{-\infty}^{-L_-}
|x-x^*|^k \ee^{-\beta_i V_0}\dd x\\
&=\frac{C}{\beta_iL_-^m}
\left|\int_{-\infty}^{-L_-}(x-x^*)^k \ee^{-\beta_i V_0}\dd x\right|.
\end{aligned}
\label{eq:app-polynomial-remainder-bound}
\end{equation}
Here $C$ depends only on the fixed polynomial coefficients and $k$, and is independent of $\beta_i$ and $L_-$. The last equality uses the fixed sign of $(x-x^*)^k$ on the integration interval. For $\beta_iE_L\gg1$, this small remainder can be absorbed into the left-hand side. Using $E_L\simeq v_mL_-^m$, we obtain
\begin{equation}
\int_{-\infty}^{-L_-}(x-x^*)^k \ee^{-\beta_i V_0}\dd x
=\frac{(-L_--x^*)^k \ee^{-\beta_i E_L}}{\beta_i F_L}
\left[1+O\!\left((\beta_i E_L)^{-1}\right)\right].
\label{eq:app-polynomial-moment-expansion}
\end{equation}
Keeping the leading term yields Eq.~\eqref{eq:junction-cancellation}.\\

Finally, we evaluate the three pieces of the last integral in Eq.~\eqref{eq:exact-left-energy-relation}, keeping the first two terms of that equation initially unchanged. Equation~\eqref{eq:tail-estimates} gives the excess-energy and constant contributions; the derivative contribution follows exactly from
$V'\ee^{-\beta_i V}=-\beta_i^{-1}\partial_x\ee^{-\beta_i V}$:
\begin{equation}
\begin{aligned}
-\beta_i(m-q)\int_{-\infty}^{-L_-}(V-E_L)\ee^{-\beta_i V}\dd x
&\simeq-\frac{m-q}{\beta_i F_L}\ee^{-\beta_i E_L},\\
-\beta_i(mE_L-F_LL_-)\int_{-\infty}^{-L_-}\ee^{-\beta_i V}\dd x
&\simeq-\frac{mE_L-F_LL_-}{F_L}\ee^{-\beta_i E_L},\\
-\beta_i\int_{-\infty}^{-L_-}x^*V'\ee^{-\beta_i V}\dd x
&=x^*[\ee^{-\beta_i V}]_{-\infty}^{-L_-}
=x^*\ee^{-\beta_i E_L}.
\end{aligned}
\label{eq:app-three-tail-pieces}
\end{equation}
The first line is the tail term retained in Eq.~\eqref{eq:left-energy-relation}. To combine the other two lines, we use the matching condition~\eqref{eq:junction-matching-identity} and then the polynomial-tail estimate~\eqref{eq:junction-cancellation}:
\begin{equation}
\begin{aligned}
-\frac{mE_L-F_LL_-}{F_L}e^{-\beta_i E_L}+x^*e^{-\beta_i E_L}
&=-\frac{mE_L-F_L(L_-+x^*)}{F_L}e^{-\beta_i E_L}\\
&=-\sum_{k=0}^{m-1}(m-k)v_k(-L_--x^*)^k
\frac{e^{-\beta_i E_L}}{F_L}\\
&\simeq-\beta_i\sum_{k=0}^{m-1}(m-k)v_k
\int_{-\infty}^{-L_-}(x-x^*)^ke^{-\beta_i V_0}dx.
\end{aligned}
\label{eq:app-combined-junction-pieces}
\end{equation}
This supplies the missing part of the core moments in the second term of Eq.~\eqref{eq:exact-left-energy-relation}: the intervals $(-\infty,-L_-)$ and $[-L_-,x^*]$ join to $(-\infty,x^*]$. The term $m\beta_i U_-Zp_-$ is unchanged. This gives Eq.~\eqref{eq:left-energy-relation}. All approximate equalities retain the leading remote-junction contributions for fixed $q\neq m$; $Zp_-$ and $U_-$ include the actual tail potential throughout.

\subsection{Eigenfunction and projection bounds}
\label{sec:bound-details}

We now consider either tail $j=L,R$ and let $s\ge0$ denote the outward distance from the corresponding junction, where $V=E_j$ and $\ell_2=B_j$. For $q\ge1$, $\dd V/\dd s\ge F_j$ over the region where the tail develops. The backward equation in this coordinate is $\beta^{-1}\dd^2\ell_2/\dd s^2-(\dd V/\dd s)\dd\ell_2/\dd s=-\lambda_2\ell_2$. Replacing $\dd V/\dd s$ with $F_j$ gives the admissible solution $B_j\ee^{rs}$, where, for $F_j^2>4\lambda_2/\beta$,
\begin{equation}
r=\frac{\beta}{2}
\left(F_j-\sqrt{F_j^2-4\lambda_2/\beta}\right)
=\frac{2\lambda_2}{F_j+\sqrt{F_j^2-4\lambda_2/\beta}}
\le\frac{2\lambda_2}{F_j}.
\label{eq:app-constant-force-exponent}
\end{equation}
The unique node of $\ell_2$ lies between the junctions, so $u(s)=\ell_2(s)/B_j>0$ on either tail, with $u(0)=1$. Primes below denote derivatives with respect to $s$. The backward equation and $r^2-\beta F_jr+\beta\lambda_2=0$ give
\begin{equation}
\frac{\dd}{\dd s}\left[\ee^{-\beta V+rs}(u'-ru)\right]
=\beta r(V'-F_j)\ee^{-\beta V+rs}u\ge0.
\label{eq:app-tail-comparison-identity}
\end{equation}
Applying the Cauchy--Schwarz inequality gives
\begin{equation}
\begin{aligned}
\int_0^\infty\left|\ee^{-\beta V+rs}(u'-ru)\right|\dd s
&=\int_0^\infty
\left(\ee^{-\beta V/2+rs}\right)
\left(\ee^{-\beta V/2}|u'-ru|\right)\dd s\\
&\le\left(\int_0^\infty\ee^{-\beta V+2rs}\dd s\right)^{1/2}
\left(\int_0^\infty(u'-ru)^2\ee^{-\beta V}\dd s\right)^{1/2}\\
&\le\left(\frac{\ee^{-\beta E_j}}{\beta F_j-2r}\right)^{1/2}
\left(2\int_0^\infty[u'^2+r^2u^2]\ee^{-\beta V}\dd s\right)^{1/2}
<\infty.
\end{aligned}
\label{eq:app-tail-comparison-integrability}
\end{equation}
The last line uses $V(s)\ge E_j+F_js$, $2r<\beta F_j$, and $(u'-ru)^2\le2u'^2+2r^2u^2$. The biorthogonal normalization and the backward eigenvalue equation ensure that both $u$ and $u'$ are square integrable with the Boltzmann weight $\ee^{-\beta V}$ on each tail. The bracketed quantity in Eq.~\eqref{eq:app-tail-comparison-identity} is therefore integrable and nondecreasing, so it tends to zero at infinity. Hence $u'\le ru$. Integrating the backward equation from $s$ to infinity with the natural boundary condition gives $u'\ge0$, which, together with $u'\le ru$, $u(0)=1$, and $r\le2\lambda_2/F_j$, yields
\begin{equation}
1\le\frac{\ell_2(s)}{B_j}\le \ee^{rs},
\qquad
|\ell_2(s)-B_j|
\le |B_j|\left(\ee^{2\lambda_2s/F_j}-1\right).
\label{eq:app-tail-mode-comparison}
\end{equation}
This applies to both signs of $B_j$.
Finally, integrating $\dd V/\dd s\ge F_j$ gives $s\le(V-E_j)/F_j$. Using $\ee^z-1\le z\ee^z$ for $z\ge0$, we obtain
\begin{equation}
\begin{aligned}
|\ell_2-B_j|
&\le |B_j|\left\{\exp\!\left[\frac{2\lambda_2(V-E_j)}{F_j^2}\right]-1\right\}\\
&\le |B_j|\frac{2\lambda_2(V-E_j)}{F_j^2}
\exp\!\left[\frac{2\lambda_2(V-E_j)}{F_j^2}\right],
\end{aligned}
\label{eq:app-tail-mode-energy-bound}
\end{equation}
which is Eq.~\eqref{eq:tail-mode-bound}.

Taking the absolute value in Eq.~\eqref{eq:tail-projection-derivative} gives
\begin{equation}
\left|\frac{\partial a_2^{(1)}}{\partial\beta_i}\right|
\le\frac1Z\sum_{j=L,R}\int_{\mathrm{tail}\,j}
|U_i-V|\,|\ell_2-B_j|\ee^{-\beta_i V}\dd x.
\label{eq:app-absolute-projection-derivative}
\end{equation}
We now set $u=\beta_i(V-E_j)\ge0$ and bound the three factors separately.\\

For the eigenfunction correction and Boltzmann factor, Eq.~\eqref{eq:tail-mode-bound} gives
\begin{equation}
\begin{aligned}
|\ell_2-B_j|\ee^{-\beta_i V}
&\le\frac{2\lambda_2|B_j|}{\beta_i F_j^2}
u\ee^{-\beta_i E_j}
\exp\!\left[-\left(1-\frac{2\lambda_2}{\beta_i F_j^2}\right)u\right]\\
&\le\frac{2\lambda_2|B_j|}{\beta_i F_j^2}
u\ee^{-\beta_i E_j}\ee^{-u/2},
\end{aligned}
\label{eq:app-weighted-mode-bound}
\end{equation}
where the second inequality uses $2\lambda_2/(\beta_i F_j^2)\le1/2$.

As $\beta_i\to0$ with $\beta_i E_{L,R}\gg1$, the tail contributions are exponentially suppressed and the leading core potential is $v_m(x-x^*)^m$. Rescaling $x-x^*$ by $(\beta_i v_m)^{-1/m}$ gives $Z\propto\beta_i^{-1/m}$ to leading order. Applying the same rescaling to the energy integral gives
\begin{equation}
U_i=-\frac{\partial\ln Z}{\partial\beta_i}
=\frac{1/m+o(1)}{\beta_i},
\qquad |U_i|\le\frac1{\beta_i},
\end{equation}
where the coefficient $1/m$ follows by integration by parts, and the bound holds for sufficiently remote junctions since $m\ge4$.\\

For the energy difference, the definition $u=\beta_i(V-E_j)$ gives the exact identity $V=E_j+u/\beta_i$. Since $E_j>0$ and $u\ge0$, the triangle inequality gives
\begin{equation}
|U_i-V|\le |U_i|+V
\le E_j+\frac{1+u}{\beta_i}
\le(E_j+\beta_i^{-1})(1+u).
\end{equation}
The last inequality adds the nonnegative term $E_ju$.\\

Multiplying the three bounds yields
\begin{equation}
\left|\frac{\partial a_2^{(1)}}{\partial\beta_i}\right|
\le\frac{2\lambda_2}{Z}\sum_{j=L,R}
\frac{|B_j|\ee^{-\beta_i E_j}}{\beta_i^2F_j^3}
(E_j+\beta_i^{-1})\int_0^\infty z(1+z)\ee^{-z/2}\dd z,
\label{eq:app-projection-derivative-bound}
\end{equation}
which is Eq.~\eqref{eq:dimensionless-tail-bound}. The remaining integral equals $20$, giving the numerical factor $40$ in Eq.~\eqref{eq:scaled-tail-bound}.

\bibliography{mpemba-gemini}